\documentclass[twocolumn]{aastex63}

\usepackage[T1]{fontenc}

\usepackage{amsmath}

\newcommand{\package}[1]{\textsl{#1}}
\newcommand{\gaia}{\textsl{Gaia}}
\newcommand{\gyr}{\ensuremath{\textrm{Gyr}}}

\newcommand{\ul}{\ensuremath{\textrm{kpc}^2\,\textrm{Myr}^{-1}}}
\newcommand{\ue}{\ensuremath{\textrm{kpc}^2\,\textrm{Myr}^{-2}}}

\newcommand{\feh}{\ensuremath{\textrm{[Fe/H]}}}
\newcommand{\afe}{\ensuremath{\textrm{[$\alpha$/Fe]}}}

\shorttitle{}
\shortauthors{bonaca et al.}

\begin{document}\sloppy\sloppypar\raggedbottom\frenchspacing 

\title{Orbital Clustering Identifies the Origins of Galactic Stellar Streams}

\correspondingauthor{Ana~Bonaca}
\email{ana.bonaca@cfa.harvard.edu}

\author[0000-0002-7846-9787]{Ana~Bonaca}
\affil{Center for Astrophysics | Harvard \& Smithsonian, 60 Garden Street, Cambridge, MA 02138, USA}

\author[0000-0003-3997-5705]{Rohan~P.~Naidu}
\affil{Center for Astrophysics | Harvard \& Smithsonian, 60 Garden Street, Cambridge, MA 02138, USA}

\author[0000-0002-1590-8551]{Charlie~Conroy}
\affil{Center for Astrophysics | Harvard \& Smithsonian, 60 Garden Street, Cambridge, MA 02138, USA}

\author[0000-0003-2352-3202]{Nelson~Caldwell}
\affil{Center for Astrophysics | Harvard \& Smithsonian, 60 Garden Street, Cambridge, MA 02138, USA}

\author[0000-0002-1617-8917]{Phillip~A.~Cargile}
\affil{Center for Astrophysics | Harvard \& Smithsonian, 60 Garden Street, Cambridge, MA 02138, USA}

\author[0000-0002-6800-5778]{Jiwon~Jesse~Han}
\affil{Center for Astrophysics | Harvard \& Smithsonian, 60 Garden Street, Cambridge, MA 02138, USA}

\author[0000-0002-9280-7594]{Benjamin~D.~Johnson}
\affil{Center for Astrophysics | Harvard \& Smithsonian, 60 Garden Street, Cambridge, MA 02138, USA}

\author[0000-0002-8804-0212]{J.~M.~Diederik~Kruijssen}
\affiliation{Astronomisches Rechen-Institut, Zentrum f\" ur Astronomie der Universit\" at Heidelberg, M\" onchhofstra\ss e 12-14, D-69120 Heidelberg, Germany}

\author[0000-0002-5629-8876]{G.~C.~Myeong}
\affil{Center for Astrophysics | Harvard \& Smithsonian, 60 Garden Street, Cambridge, MA 02138, USA}

\author[0000-0003-2573-9832]{Josh~Speagle}
\affiliation{University of Toronto, Department of Statistical Sciences, Toronto, M5S 3G3, Canada}
\affiliation{University of Toronto, David A. Dunlap Department of Astronomy \& Astrophysics, Toronto, M5S 3H4, Canada}
\affiliation{University of Toronto, Dunlap Institute for Astronomy \& Astrophysics, Toronto, M5S 3H4, Canada}

\author[0000-0001-5082-9536]{Yuan-Sen~Ting}
\affil{Institute for Advanced Study, Princeton, NJ 08540, USA}
\affil{Department of Astrophysical Sciences, Princeton University, Princeton, NJ 08544, USA}
\affil{Observatories of the Carnegie Institution of Washington, 813 Santa Barbara Street, Pasadena, CA 91101, USA}
\affil{Research School of Astronomy and Astrophysics, Mount Stromlo Observatory, Cotter Road, Weston Creek, ACT 2611, Canberra, Australia}

\author[0000-0002-5177-727X]{Dennis~Zaritsky}
\affil{Steward Observatory and University of Arizona, 933 N. Cherry Ave, Tucson, AZ 85719, USA}

\begin{abstract}\noindent 
The origins of most stellar streams in the Milky Way are unknown.
With improved proper motions provided by \gaia\ EDR3, we show that the orbits of 23 Galactic stellar streams are highly clustered in orbital phase space.
Based on their energies and angular momenta, most streams in our sample can plausibly be associated with a specific (disrupted) dwarf galaxy host that brought them into the Milky Way.
For eight streams we also identify likely globular cluster progenitors (four of these associations are reported here for the first time).
Some of these stream progenitors are surprisingly far apart, displaced from their tidal debris by a few to tens of degrees.
We identify stellar streams that appear spatially distinct, but whose similar orbits indicate they likely originate from the same progenitor.
If confirmed as physical discontinuities, they will provide strong constraints on the mass-loss from the progenitor.
The nearly universal ex-situ origin of existing stellar streams makes them valuable tracers of galaxy mergers and dynamical friction within the Galactic halo.
Their phase-space clustering can be leveraged to construct a precise global map of dark matter in the Milky Way, while their internal structure may hold clues to the small-scale structure of dark matter in their original host galaxies.
\end{abstract}

\section{Introduction}
\label{sec:intro}

Stellar streams hold the promise of delivering fundamental insights about the Galaxy and the nature of dark matter.
Due to their cold kinematics, even subtle gravitational perturbations will leave a prominent observational signature \citep[e.g., gaps due to encounters with dark matter subhalos, fans due to encounters with the bar,][]{johnston2002, pearson2017, bonaca2019a}.
A remarkable and unexpected discovery over the past few years is that almost every stream displays a variety of complex morphologies, large velocity dispersions, and/or widths incommensurate with dynamically unperturbed models \citep[e.g.,][]{pwb, bonaca2019b, bonaca2020a, li2020, gialluca2020}.

A key missing piece of context in our modeling of streams is knowledge of their origins. 
We have thus far been unable to answer some of the most basic of questions: were the stream progenitors born in-situ, or did they enter the Milky Way during mergers with dwarf galaxies?
Which streams originate from disrupting globular clusters, and which from dissolving dwarf galaxies?
Understanding the origin of streams is critical to fulfilling their promise as probes of fundamental physics and the nature of the Galaxy.
For instance, properties of the progenitor dwarf galaxy, including its mass and density profile, will leave imprints on stream structure and velocity dispersion \citep[][]{carlberg2018, malhan2020}.
Other aspects of the dwarf host (e.g., its accretion redshift) can help bracket for how long the stream must be orbiting the Milky Way, and thus constrain important parameters such as the expected number of subhalo encounters \citep[e.g.,][]{erkal2016}.

In this paper we use stream proper motions from \gaia\ and ground-based radial velocities to derive orbits, including angular momenta and energies.
This information, along with metallicities where available, is used to associate 23 stellar streams with disrupted dwarf galaxies and/or known globular clusters.
Throughout the paper we use the term ``stream progenitor'' for an object whose tidal disruption formed that stream and the term ``host galaxy'' for a galaxy in which the stream progenitor was born.

\section{Stream Orbits}
\label{sec:streamorbits}
More than 60 long and thin stellar streams have been discovered in the Milky Way \citep[see][for an up-to-date catalog]{mateu2018}.
Due to their small width, these are expected to be tidally dissolved globular clusters, although low-mass dwarf galaxy progenitors are also allowed (and can be distinguished with the presence of a metallicity spread).
Only a handful of streams directly connect to a surviving globular cluster \citep[e.g.,][]{rockosi2002, grillmair2006b}.
The vast majority of streams have no apparent progenitor within the stream and here we explore the origin of such streams.

Thanks to data released by the \gaia\ mission, 3D positions and two proper motion components are known for a sample of 23 streams without a discernible progenitor \citep{ibata2019, shipp2019, riley2020}.
\gaia\ DR2 proper motions detected in Elqui, Phoenix, Turbio, Turranburra, and Willka Yaku are significantly more uncertain than along other streams in our sample.
For these streams we selected blue horizontal branch stars from the \gaia\ EDR3 catalog \citep{gaiaedr3} with a box in the \gaia\ color-magnitude diagram at the expected stream distance, and used their more precise proper motions in orbit fitting.
Radial velocities have been measured for five of these streams, thereby providing full 6D phase-space information \citep{caldwell2020, li2020, bonaca2020b}.

In this work, we use stream orbits derived by Bonaca \& Kruijssen (submitted), and provide here a brief overview of their fitting procedure.
Assuming a static, axisymmetric model of the Milky Way \citep[v1.2 \texttt{MilkyWayPotential}]{gala}, Bonaca \& Kruijssen used the available stream data to constrain their orbits.
They sampled the stream orbital parameters using a Monte Carlo Markov Chain ensemble sampler and provide direct samples from the posterior to account for correlations between parameters.
In this work we characterize the orbit by its total energy, $E_{\rm tot}$, and the $z$ component of the angular momentum, $L_z$ (perpendicular to the Galactic disk)---both conserved quantities in the adopted gravitational potential.
We further employ the orthogonal component of the angular momentum $L_\perp \equiv \sqrt{L_x^2 + L_y^2}$, which is not fully conserved in an axisymmetric potential, but is still a useful quantity for orbital characterization \citep[e.g.,][]{helmi1999}.
We use a right-handed coordinate system, such that $L_z<0$ denotes prograde orbits.

\begin{figure*}
\begin{center}
\includegraphics[width=\textwidth]{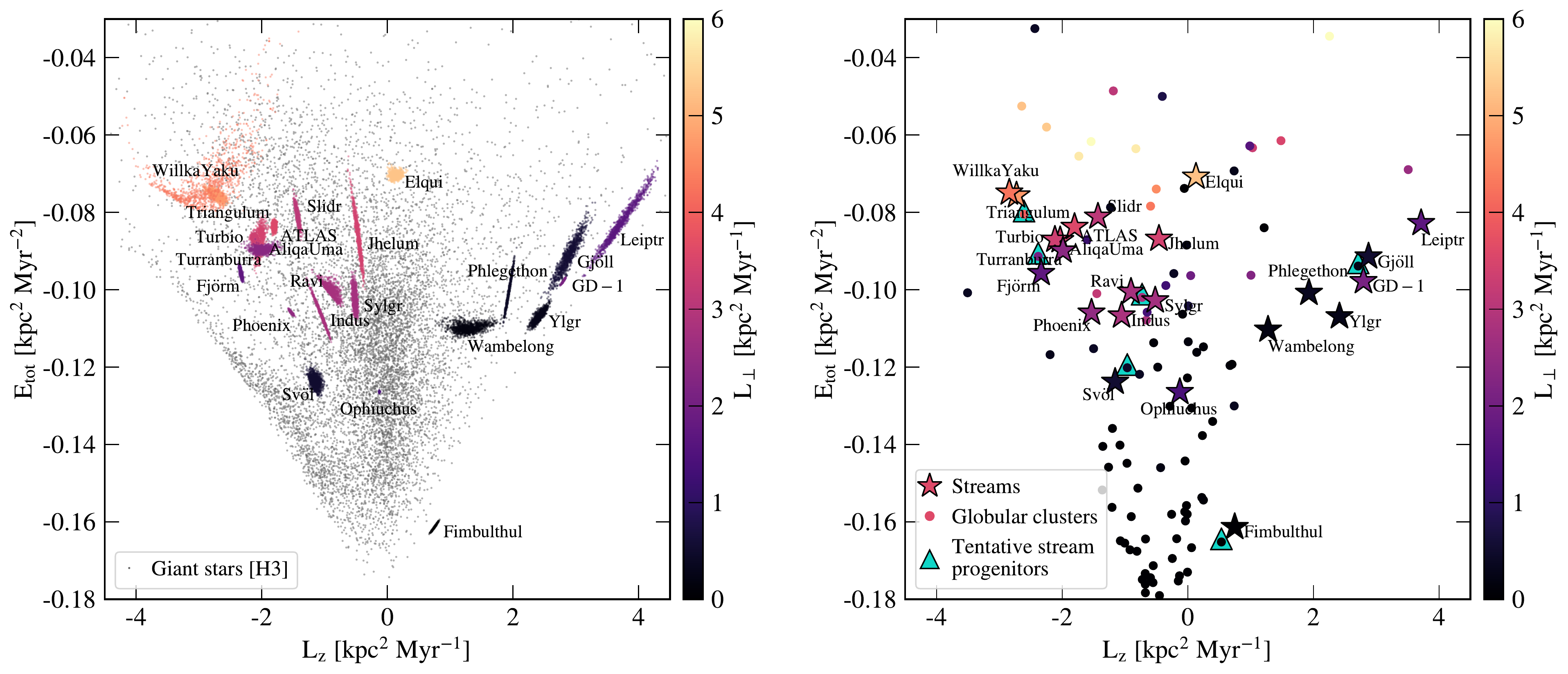}
\end{center}
\caption{
Left: Posterior samples of orbital energy and angular momentum $L_z$ for stellar streams, colored by the average orthogonal component of the angular momentum, $L_\perp$, and compared to field stars (black points).
Right: Median energy and angular momentum for stellar streams (stars) compared to globular clusters (circles), both color-coded by $L_\perp$.
Unlike stars and clusters, streams predominantly occupy tangential orbits (large $|L_z|$) and are more strongly clustered in phase space.
Globular clusters in close proximity to streams are indicated as plausible progenitors (cyan triangles).
}
\label{fig:elz}
\end{figure*}

\section{Streams in phase space}
\label{sec:phasespace}

\subsection{Overview}
\label{sec:elz}

We present the phase-space distribution of Galactic streams in Figure~\ref{fig:elz}.
In the left panel, each stream is represented in energy and $L_z$ angular momentum with 1000 samples from the posterior distributions, while the medians of these distributions are shown as stars in the right panel.
In both panels points are color-coded by the in-plane component of the angular momentum, $L_\perp$.
As a comparison, we include the phase-space distribution of giant stars from the H3 Spectroscopic Survey \citep[left panel, small black points;][]{conroy2019}, and Galactic globular clusters \citep[right panel, small circles colored by $L_\perp$;][]{baumgardt2019}.

The two most striking features of streams in phase space are: (1) the significant degree of clustering, and (2) the lack of streams on radial orbits ($L_z\approx0$).
In contrast, the major feature in stars and globular clusters is the large population of objects on radial orbits, identified as debris from the Gaia-Sausage-Enceladus (GSE) merger \citep[e.g.,][]{belokurov2018, helmi2018, naidu2020}.
Only two stellar streams are found on radial orbits, Ophiuchus and Elqui.
Unlike stars and globular clusters, the majority of the streams are highly clustered in two groups, a retrograde group containing 7 streams, and a prograde group comprised of 14 streams.

Most of the retrograde streams are found in a narrow locus from $(E_{\rm tot}, L_z)\approx(-0.08\,\ue,\allowbreak 4\,\ul)$ to $(E_{\rm tot}, L_z)\approx(-0.11\,\ue,\allowbreak 1\,\ul)$, which includes Leiptr, Gj\" oll, GD-1, Phlegethon, Ylgr, and Wambelong.
Fimbulthul is on the extension of this diagonal to lower energies.
With the exception of GD-1 and Leiptr, all retrograde streams have uniformly low $L_\perp\lesssim1\,\ul$.
This clustering in orbital poles suggests that the entire retrograde group may share a common origin.
Within this group, Wambelong, Leiptr, and Gj\" oll are highly aligned, defining a potential ``plane of streams''.

The prograde group of streams appears to separate into three distinct regions: (1) Triangulum and Willka Yaku at $L_z\approx-3\,\ul$, (2) Slidr, ATLAS, Aliqa Uma, Turbio, Turranburra, and Fj\" orm at $L_z\approx-2\,\ul$, and (3) Jhelum, Sylgr, Ravi, Indus, Phoenix, and Sv\" ol at $L_z\approx-1\,\ul$.
The in-plane angular momentum decreases with decreasing $|L_z|$ and is approximately uniform within each small cluster.
The median energy of these sub-groups also decreases with decreasing $|L_z|$, while its dispersion increases, such that the most radial clump spans the largest range in energy levels.
Clustering of the prograde streams  provides tantalizing evidence of a common origin among stellar streams, which we explore further in \S\,\ref{sec:hosts}.

Several streams are also very closely associated with a globular cluster in phase space, suggesting a possible physical connection.
Clusters that appear close to streams in energy, both components of the angular momentum, and also distance and metallicity are outlined in cyan triangles in the right panel of Figure~\ref{fig:elz}.
The identified cluster--stream groups include: Omega Cen--Fimbulthul; NGC~3201--Gj\" oll; NGC~4590--Fjorm; NGC~5024--Sylgr and Ravi; NGC~5272--Sv\" ol; and NGC~5824--Triangulum.
We explore these connections further in \S\,\ref{sec:progenitors}.

\begin{figure*}
\begin{center}
\includegraphics[width=\textwidth]{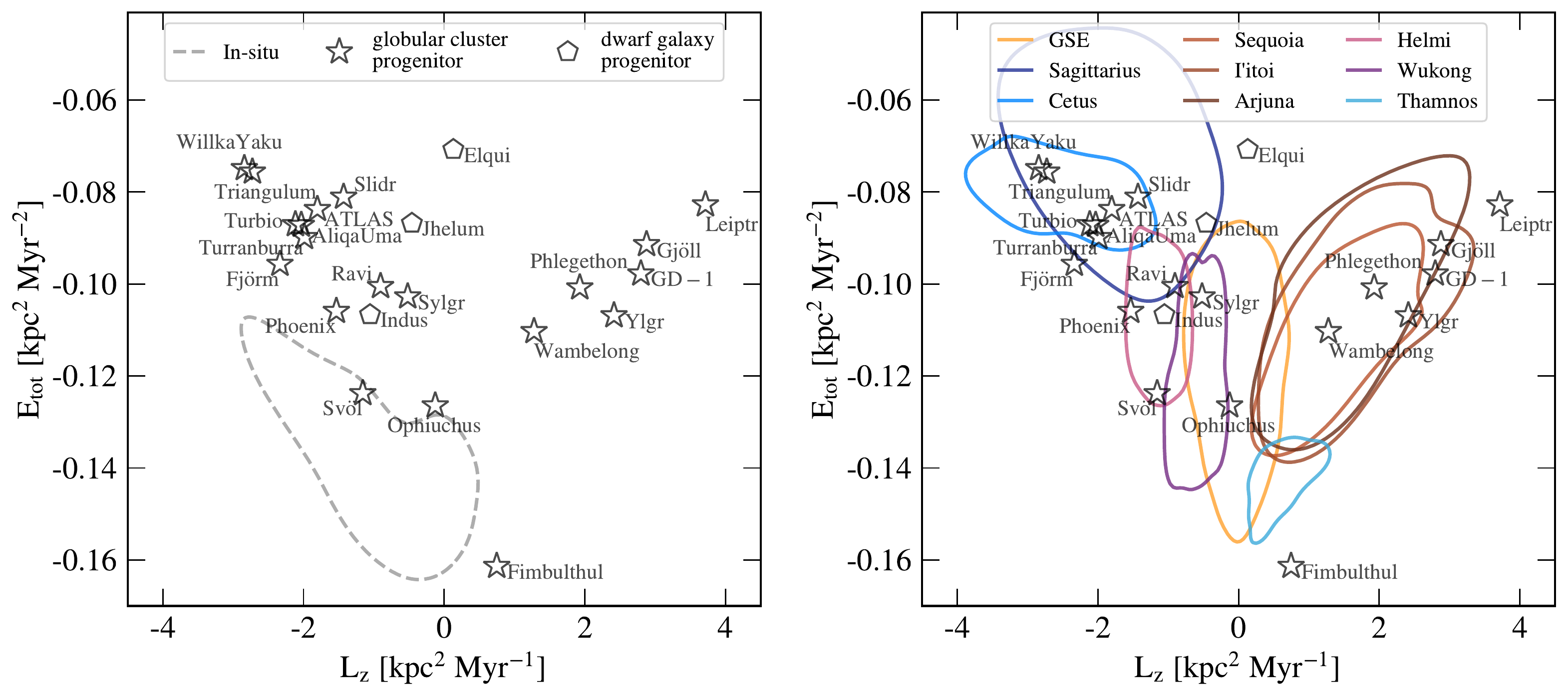}
\end{center}
\caption{
Left: The phase-space positions of stellar streams in our sample (stars) are largely outside the region of in-situ halo stars (dashed contour).
Right: Streams can be mapped to individual accretion events onto the Milky Way (contours of different colors) based on their orbits.
}
\label{fig:hosts}
\end{figure*}

\subsection{Association with disrupted galaxies}
\label{sec:hosts}

\citet{naidu2020} performed a detailed chemo-dynamical decomposition of stars observed in the H3 Spectroscopic Survey to identify structures in the Galactic halo.
In Figure~\ref{fig:hosts} we compare the locations of streams in $E-L_z$ space to the structures identified in Naidu et al.
The left panel compares the stream locations to the stars inferred to be born within the Milky Way (in-situ; dashed line), while the right panel compares streams to the populations of stars inferred to have an accretion origin.
In the right panel we use contours of different colors to mark the phase-space distribution of structures identified in \citet{naidu2020}, most of which likely constitute distinct accretion events.
Specifically, we performed kernel density estimation and encompass the region where the average density for a given component is higher than 20\,\% of its maximum.
Most of the streams have energies and angular momenta consistent with the distribution of one of the known halo substructures.

The high-energy group of retrograde streams is well aligned with debris from Sequoia, I'itoi, and Arjuna.
These three Milky Way progenitors differ in metallicity \citep{naidu2020}, so it might be possible to further refine the streams' association with these structures based on their metallicity.
While GD-1 \citep[spectroscopic $\feh=-2.3$,][]{bonaca2020b}, Ylgr \citep[spectroscopic $\feh=-1.9$,][]{ibata2019}, and  Wambelong \citep[isochrone $\feh=-2.2$,][]{shipp2018} have low metallicities that can be plausibly associated with any of these progenitors, Gj\" oll \citep[spectroscopic $\feh=-1.5$,][]{hansen2020}, Leiptr \citep[isochrone $\feh=-1.6$,][]{ibata2019}, and Phlegethon \citep[spectroscopic $\feh=-1.6$,][]{ibata2018} are sufficiently metal-rich to favor association with Sequoia or Arjuna.

Turning now toward the streams on more radial orbits, Fimbulthul lies at low energies associated with the inner Galaxy, beyond the region of energies and angular momenta surveyed by H3.
Despite being slightly retrograde, its high metallicity rules out association with the low-mass satellites on retrograde orbits.
On the other hand, numerical simulations show that debris from the massive GSE can attain similar angular momentum at these energies (R. Naidu et al., in preparation).
We, therefore, associate GSE as the host galaxy of Fimbulthul.

Ophiuchus is on a low-energy radial orbit and we therefore associate it with GSE.
Although some in-situ stars have orbits similar to Ophiuchus, the low metallicity of $\feh=-1.95$ \citep{sesar2015} disfavors Ophiuchus as an in-situ stellar stream.
Furthermore, the curious morphology of Ophiuchus (the short extent of the narrow, dense stream, which is surrounded by a wide, low-density envelope), are naturally expected from globular clusters that first started dissolving in a dwarf galaxy host and continued dissolving in the Milky Way upon the host galaxy's disruption (see \citealt{carlberg2018, malhan2020}).

Elqui is on a high-energy radial orbit and appears unassociated with any of the Milky Way progenitors.
\citet{ji2020} found significant metallicity spread in Elqui, a telltale signature of a dwarf galaxy origin with an extended star-formation history.
Elqui could plausibly have been a satellite of GSE, but we disfavor this association because GSE accreted $\approx10\,\gyr$ ago \citep{bonaca2020c} and disruption of satellites on radial orbits is fast.
Most likely, Elqui is tidal debris of a recently accreted dwarf galaxy without a more massive host.

\begin{deluxetable*}{l c c c c}
\tablehead{
Name & Host Galaxy Candidate & Progenitor & Associated with & Type
}
\decimals
\startdata
Aliqa Uma & Sagittarius & \dots & ATLAS & GC\\ 
ATLAS & Sagittarius & \dots & Aliqa Uma & GC\\ 
Elqui & none & itself & \dots & DG\\ 
Fimbulthul & Gaia-Sausage-Enceladus & NGC 5139 & \dots & GC\\ 
Fj\"{o}rm & Sagittarius & NGC 4590 & \dots & GC\\ 
GD-1 & Sequoia / Arjuna / I'itoi & \dots & \dots & GC\\ 
Gj\"{o}ll & Sequoia / Arjuna & NGC 3201 & \dots & GC\\ 
Indus & \dots & (Wukong / Helmi) & Jhelum & DG\\ 
Jhelum & \dots & (Wukong / Helmi) & Indus & DG\\ 
Leiptr & Sequoia / Arjuna & \dots & \dots & GC\\ 
Ophiuchus & Gaia-Sausage-Enceladus & \dots & \dots & GC\\ 
Phlegethon & Sequoia / Arjuna & \dots & \dots & GC\\ 
Phoenix & Helmi / Wukong & \dots & \dots & GC\\ 
Ravi & Helmi / Wukong & NGC 5024 & Sylgr & GC\\ 
Slidr & Sagittarius & \dots & \dots & GC\\ 
Sv\"{o}l & in situ / Helmi / Wukong & NGC 5272 & \dots & GC\\ 
Sylgr & Helmi / Wukong & NGC 5024 & Ravi & GC\\ 
Triangulum & Cetus & NGC 5824 & Turbio & GC\\ 
Turbio & Cetus & NGC 5824 & Triangulum & GC\\ 
Turranburra & Sagittarius & \dots & \dots & GC\\ 
Wambelong & Sequoia / Arjuna / I'itoi & \dots & \dots & GC\\ 
Willka Yaku & Cetus & \dots & \dots & GC\\ 
Ylgr & Sequoia / Arjuna / I'itoi & \dots & \dots & GC\\ 

\enddata
\caption{
The origins of stellar streams in the Milky Way.
Progenitor is the object that dissolved to create the stellar stream.
The last column determines the progenitor as globular cluster (GC) or dwarf galaxy (DG).
Host galaxy is the galaxy that brought the stream progenitor into the Milky Way.
Tentative host galaxy candidates and progenitors are placed in parentheses.  Unknown or very uncertain associations are labelled with ellipses.
}
\label{table}
\end{deluxetable*}

At slightly lower energies and on prograde orbits, Jhelum and Indus are two streams that have been linked together based on their 3D orbits \citep{bonaca2019b}.
Curiously, they have similar metallicities ([Fe/H]$\approx-2.1$) and detectable spread in chemical abundances \citep{ji2020}, which suggests they are tidal debris from the same dwarf galaxy at different orbital phases, possibly from separate pericenter passages.
This region of phase space is occupied by debris from Helmi\footnote{In the literature this structure is often referred to as the Helmi Streams \citep{helmi1999}. To avoid confusion with the thin streams discussed in this work, we refer to this structure simply as ``Helmi'' throughout.} and Wukong; we therefore suggest that Jhelum and Indus may represent the remaining coherent debris from either of these galaxies.
Wukong is the more metal-poor of the two ($\feh\approx-1.6$ vs. $-1.3$), which, given the low metallicities of Jhelum and Indus, would favor an association with Wukong over Helmi.
Alternatively, Jhelum and Indus might originate from a satellite of a more massive merger like Sagittarius or GSE.

Ravi, Sylgr and Phoenix are thin streams with orbits similar to Indus.
With little metallicity variation within either Sylgr or Phoenix \citep{ibata2019, wan2020}, these streams likely originate from globular clusters hosted by the progenitor of Indus and Jhelum.

Sv\" ol is the final stream in this group of low and prograde angular momentum, found at substantially lower energy from the rest of the group.
Based on its energy and angular momentum, Sv\" ol may be associated with Helmi, Wukong, or the in-situ component of the stellar halo.
Interestingly, \citet{ibata2019} identified a star with Sv\" ol kinematics, but marked it as a probable contaminant because of a metallicity $\feh=-1.08$.
If this star is confirmed to be associated with Sv\" ol and is determined to have a high \afe\ abundance, Sv\" ol would be the first known halo stream to be strongly associated with an in-situ population in the Milky Way. 

\begin{figure*}
\begin{center}
\includegraphics[width=0.9\textwidth]{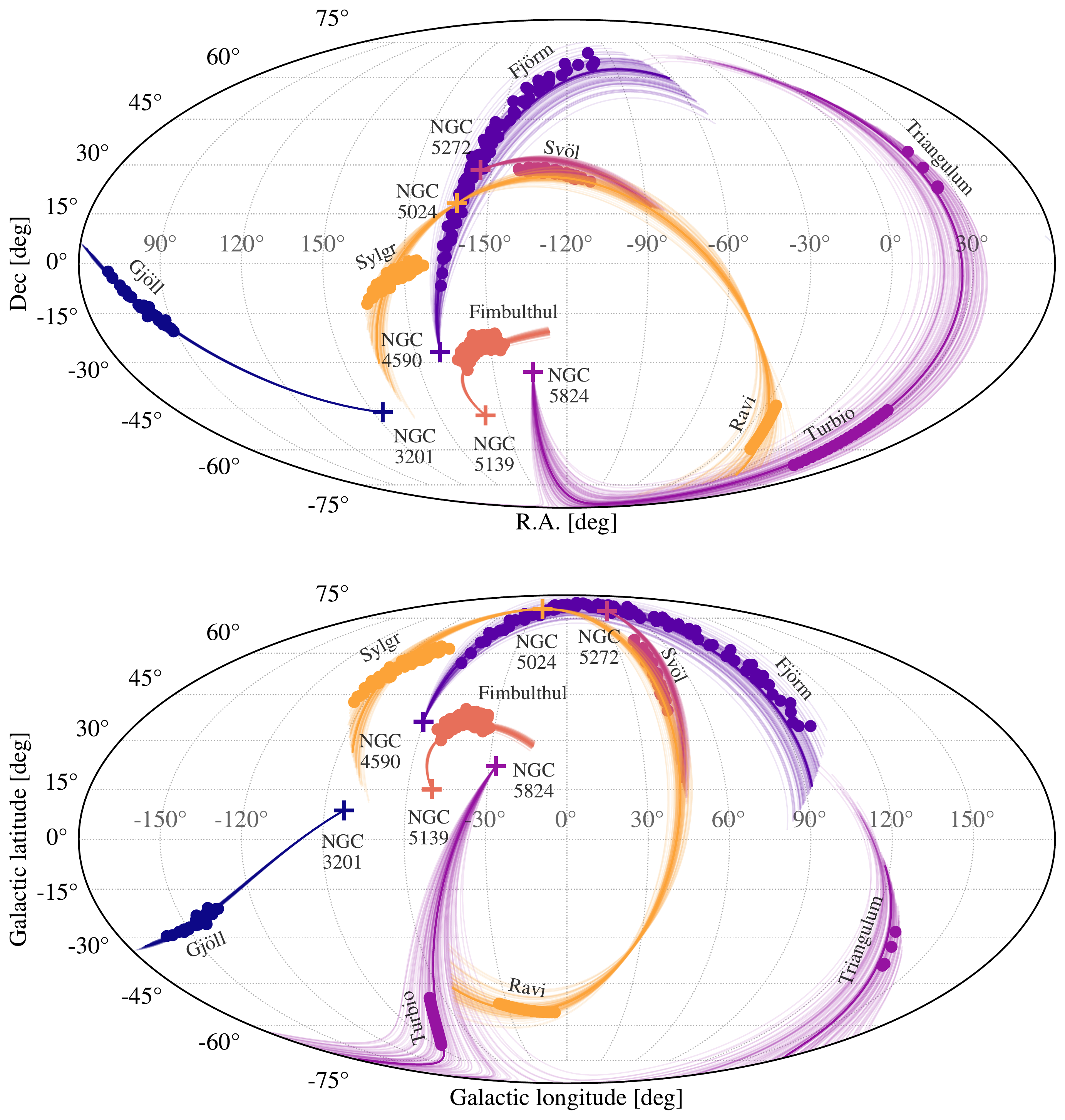}
\end{center}
\caption{
Sky positions of stellar streams (circles) and globular clusters (crosses) that have similar orbital energies and angular momenta (see Figure~\ref{fig:elz}), shown in equatorial coordinates in the top and Galactic at the bottom.
Thick lines are the best-fit orbits of globular clusters, whereas thin lines sample observational uncertainties in the clusters' 6D positions.
Despite being spatially separated, these clusters are likely stream progenitors because their orbits connect them to the streams.
}
\label{fig:sky}
\end{figure*}

A large group of streams including Aliqa Uma, ATLAS, Fj\" orm, Slidr, and Turranburra forms a tight sequence at intermediate prograde angular momenta (centered on $L_z\sim-2$ kpc$^2$ Mpc$^{-1}$).
Stars from both the Cetus stream and the Sagittarius dwarf galaxy have been observed at these $E-L_z$ (Figure~\ref{fig:hosts}).
We associate these streams with Sagittarius based on their low $L_y$ angular momenta \citep{Johnson2020}.
All of these streams are thin and no metallicity spreads have been detected, which suggests they were originally globular clusters associated with Sagittarius that were disrupted by the Milky Way tidal force.
Several streams in this group are very close in the phase space, which suggests that some of them might be part of the same stream, despite appearing spatially distinct in the sky.
In fact, \citet{li2020} found that ATLAS and Aliqa Uma have a radial velocity gradient consistent with the same orbit.
Perturbations from the dynamic Sagittarius environment can produce a large gap in the originally continuous stream \citep{bonaca2020b, deboer2020}.

Finally, Triangulum and Willka Yaku form a distinct association on highly prograde, high energy orbits.
This region of phase space is occupied by stars from the Cetus stream \citep{yuan2019}, with an extension to lower orbits that captures Turbio, a stream we find associated with Triangulum in \S\ref{sec:progenitors}.
Triangulum, Turbio, and Willka Yaku also overlap with Cetus spatially, but they are significantly narrower than the Cetus stream \citep[$0.25\,\deg$ vs. $\approx2\,\deg$,][]{bonaca2012, shipp2018, newberg2009}.
This suggests that Triangulum, Turbio, and Willka Yaku are dissolved globular clusters that were brought into the Milky Way by the progenitor of the Cetus stream.

These results imply that the original host galaxy for a large fraction of stellar streams in the Galactic halo was not the Milky Way, but one of its lower-mass progenitors.
Tentative associations of streams and their host galaxies are listed in Table~\ref{table}.

\subsection{Association with globular clusters}
\label{sec:progenitors}
In this section we explore whether globular cluster progenitors of stellar streams can be identified based on their proximity in phase space.
Inspection of the right panel of Figure~\ref{fig:elz} reveals several cases in which a globular cluster is very close in $E-L_z-L_\perp$ space to one or more stellar streams.
Figure~\ref{fig:sky} shows the sky positions of the six most compelling associations (close also in distance and metallicity), with equatorial coordinates in the top panel and Galactic in the bottom panel.
In this figure the globular clusters are shown as crosses and stream stars as circles (associated objects have the same color).
Starting with the best-fit 6D positions from \citet{baumgardt2019} we integrated orbits of these clusters and in all cases found that they clearly connect to stellar streams (shown as thick lines of matching colors in Figure~\ref{fig:sky}).
Thin lines show a 100 samples from the observational uncertainties.
The connections between NGC~5139 (Omega Cen) and Fimbulthul, NGC 3201 and Gj\" oll, and NGC~4590 (M~68) and Fj\" orm have been previously identified \citep[and references therein]{ibata2021}.
These results suggest that associations between a globular cluster and a stellar stream, where the stream does not connect directly to the cluster, are common in the Milky Way (eight out of 23 streams).

These associations challenge the established picture in which tidal tails are developed symmetrically around globular clusters through steady mass loss \citep[e.g.,][]{kuepper2010}.
None of the streams here connects directly to the progenitor and we identified both the leading and the trailing tails only for NGC~5024.
Parts of stellar streams might be missing due to observational limitations, such as footprints of photometric surveys in the case of Ravi, Triangulum, and Turbio, the varying degree to which streams stand out against the bulk of the Milky Way stars in proper motions (for Fimbulthul, Fj\" orm, Gj\" oll, Sv\" ol, and Sylgr), and crowding in the disk plane (for Gj\" oll, Turbio, and Ravi).
Dedicated searches with more precise \gaia\ EDR3 proper motions \citep{gaiaedr3} and extended photometric catalogs \citep{dey2019} along the leading and trailing sides of orbits shown in Figure~\ref{fig:sky} should determine whether the stream asymmetries and gaps are physical or a selection effect.

Orbits of globular clusters in a simple model of the Milky Way match stellar streams remarkably (Figure~\ref{fig:sky}), but there are deviations that could refine our model of the Galaxy.
At certain phases of the orbit, streams can be misaligned from the progenitor's orbit \citep{sanders2013}, which may account for the misalignment of Sylgr and Ravi with the orbit of NGC~5024 and precisely constrain the streams' formation time.
Massive satellites like Sagittarius or the Large Magellanic Cloud change orbits of globular clusters \citep{garrow2020}, which may have caused the spatial offsets of Sv\" ol and Triangulum from the orbits of their matching clusters.

\section{Discussion}
\label{sec:discussion}

In this {\it Letter} we have uncovered the origin of 23 cold stellar streams in the Milky Way halo from their clustering in the phase space of orbital energies and angular momenta.
For 20 streams we identified host galaxies that brought them into the Milky Way, and found that only Sv\" ol plausibly originated from a globular cluster born in-situ.
We used the proximity of stellar streams in phase space to identify streams that appear as separate entities spatially, but which have a common orbit and are therefore part of a single, much more extended structure.
Finally, we identified six globular clusters as progenitors of eight stellar streams (the orbits of NGC~5024 and NGC~5824 each passes through two streams).
The host galaxies and individual stream progenitors are summarized in Table~\ref{table}.

The extragalactic origin of stellar streams may provide new insight into low-mass galaxies.
The progenitors of these streams are globular clusters that have until recently been a part of their host galaxies' globular cluster system.
This recently dissolved population might shed light on the origin of scatter in the relation between the globular cluster system and the galaxy mass for low-mass galaxies \citep[e.g.,][]{harris2013}.
On the other hand, assuming that streams were formed only recently would suggest larger pre-infall halo masses of their host galaxies, many of which already have a sizeable population of globular clusters (e.g., $5-7$ in Sagittarius, \citealt{Johnson2020}; up to 6 in Sequoia, \citealt{myeong2019}).
This is especially pertinent for the accretion of Sagittarius as its impact throughout the Milky Way strongly depends on its mass \citep{laporte2019}.

Our findings have wide implications on stellar streams as tracers of dark matter.
To list a few:
(1) associating shorter streams into a single, longer structure increases their sensitivity to global properties of the gravitational potential \citep{bh2018};
(2) high clustering of many stellar streams in phase space can be directly used to constrain the gravitational potential \citep{sanderson2015, reino2020};
(3) if the retrograde streams represent debris from a single merger event, then they depict dynamical friction in action \citep[e.g.,][]{chandrasekhar1942, white1978}, whose magnitude depends on the nature of dark matter \citep[e.g.,][]{lancaster2020};
(4) a known stream progenitor allows the construction of direct N-body models that best capture the inherent structure of stellar streams \citep[e.g.,][]{kuepper2008, just2009} and improve their modeling in the Milky Way potential \citep{kuepper2015};
(5) a known host galaxy provides the time of accretion onto the Milky Way \citep[e.g.,][]{kruijssen2020}, which allows for properly capturing a stream's early formation in its host galaxy \citep[e.g.,][]{carlberg2018, malhan2020} as well as subsequent evolution in the Milky Way that accurately accounts for perturbations from molecular clouds, spiral arms, and dark-matter subhalos \citep[e.g.,][]{erkal2016, banik2019}.

We close by discussing new stream observations that are most urgently needed.
Expanding this study to the entire population of streams in the Milky Way requires high quality proper motions of $\gtrsim40$ streams, many of which are distant and faint \citep[e.g.,][]{grillmair2009, grillmair2017}.
As we approach a complete census, understanding the selection function of streams in the Galaxy will be crucial for interpreting the phase-space distribution of streams.
For example, a selection function is needed to assess whether there is a genuine lack of streams on radial orbits \citep[as shell rather than stream morphologies are expected on such orbits,][]{hendel2015} or whether genuine radial streams are misclassified as dwarf galaxies \citep[that have anomalously large velocity dispersions and/or distance gradients,][]{kuepper2017}.
Precise radial velocities, such as those provided by the H3 \citep{conroy2019} and S5 \citep{li2019} surveys, are needed to further constrain the streams' orbits.
Metallicity and multi-element abundances are the ultimate tool in discriminating globular cluster vs. dwarf galaxy progenitor systems \citep[e.g.,][]{hansen2020}, with Sv\" ol, Indus, Jhelum, and the retrograde debris being the top priority.

\vspace{0.5cm}

\software{
\package{Astropy} \citep{astropy, astropy:2018},
\package{gala} \citep{gala},
\package{IPython} \citep{ipython},
\package{matplotlib} \citep{mpl},
\package{numpy} \citep{numpy},
\package{scipy} \citep{scipy}
}

\bibliographystyle{aasjournal}
\bibliography{elz}

\end{document}